\documentclass[sigconf]{acmart}
\usepackage{algpseudocode}
\usepackage[linesnumbered,ruled]{algorithm2e}
\usepackage{multirow}

\AtBeginDocument{%
  }

\copyrightyear{2023} 
\acmYear{2023} 
\setcopyright{acmlicensed}
\acmConference[MMAsia '23 Workshops]{ACM Multimedia Asia Workshops}{December 6--8, 2023}{Tainan, Taiwan}
\acmBooktitle{ACM Multimedia Asia Workshops (MMAsia '23 Workshops), December 6--8, 2023, Tainan, Taiwan}
\acmPrice{15.00}
\acmDOI{10.1145/3611380.3628555}
\acmISBN{979-8-4007-0326-3/23/12}

\begin{document}

\title{PhasePerturbation: Speech Data Augmentation via Phase Perturbation for Automatic Speech Recognition}

\author{Chengxi Lei}
\authornote{Chengxi Lei and Ruili Wang are also affiliated to School of Mathematical and Computational Sciences, Massey University, New Zealand.}
\email{c.lei@massey.ac.nz}
\affiliation{%
  \institution{School of Artificial Intelligence,\\ Dalian Maritime University,}
  \country{China}
}

\author{Satwinder Singh}
\authornote{Corresponding authors: Satwinder Singh and Ruili Wang}
\email{s.singh4@massey.ac.nz}
\affiliation{%
  \institution{School of Mathematical and Computational Sciences,\\ Massey University,}
  \country{New Zealand}}

\author{Feng Hou}
\email{f.hou@massey.ac.nz}
\affiliation{%
  \institution{School of Mathematical and Computational Sciences,\\ Massey University,}
  \country{New Zealand}}

\author{Xiaoyun Jia}
\email{dr.sophiajia@outlook.com}
\affiliation{%
  \institution{Institute of Governance \& School of Political Science and Public Administration,\\ Shandong University,}
  \country{China}}
  
\author{Ruili Wang}
\authornotemark[1]
\authornotemark[2]

\email{ruili.wang@massey.ac.nz}
\affiliation{%
   \institution{School of Artificial Intelligence,\\ Dalian Maritime University,}
  \country{China}
  }


\begin{abstract}
  Most of the current speech data augmentation methods operate on either the raw waveform or the amplitude spectrum of speech. In this paper, we propose a novel speech data augmentation method called PhasePerturbation that operates dynamically on the phase spectrum of speech. Instead of statically rotating a phase by a constant degree, PhasePerturbation utilizes three dynamic phase spectrum operations, i.e., a randomization operation, a frequency masking operation, and a temporal masking operation, to enhance the diversity of speech data. We conduct experiments on wav2vec2.0 pre-trained ASR models by fine-tuning them with the PhasePerturbation augmented TIMIT corpus. The experimental results demonstrate 10.9\% relative reduction in the word error rate (WER) compared with the baseline model fine-tuned without any augmentation operation. Furthermore, the proposed method achieves additional improvements (12.9\% and 15.9\%) in WER by complementing the Vocal Tract Length Perturbation (VTLP) and the SpecAug, which are both amplitude spectrum-based augmentation methods. The results highlight the capability of PhasePerturbation to improve the current amplitude spectrum-based augmentation methods.
\end{abstract}





\maketitle

\section{Introduction}

Deep learning has demonstrated remarkable effectiveness in the field of Automatic Speech Recognition (ASR) \cite{hinton_deep_2012}. Extensive research efforts have been devoted to improving ASR model network architectures, leading to significant progress \cite{dahl_context-dependent_2012,graves_speech_2013,hsu_meta_2020,winata_learning_2020,singh_improved_2022}. These ASR models often require extensive volumes of training data. However, over 80\% of languages are categorized as low-resource languages, lacking such an amount of labeled data \cite{singh_enhancing_2023,chiu_state---art_2018}. To address this issue, data augmentation methods are used to generate synthetic speech data. There are mainly two approaches for data augmentation for ASR: (i) Manipulating raw waveforms to synthesize new data, such as introducing noise into clean audio to create noisy samples \cite{hannun_deep_2014,kim_generation_2017-1,sainath_factored_2016}; (ii) Applying data augmentation directly to Mel-spectrum, which is an amplitude spectrum. These Mel-spectrum based methods treat the amplitude spectrum as an image and use the corresponding data augmentation approaches \cite{ko_audio_2015,jaitly_vocal_2013,chan_specaugment_2019,wang_specaugment_2021,song_specswap_2020,kim_specmix_2021,meng_mixspeech_2021,zhang_mixup_2018,yun_cutmix_2019,wang_cyclicaugment_2022}.

Lately, there has been a growing interest in utilizing phase information to improve Generative Adversarial Network (GAN)-based vocoders used in speech synthesis. Morrison et al. \cite{morrison_chunked_2022} have shown that vocoder generators could reliably rebuild signal phases from the Mel-spectrum, even without the initial phase. PhaseAug \cite{lee_phaseaug_2023} was proposed to address the overfitting of the discriminator. The proposed approach rotated the phase in the frequency domain to simulate one-to-many mapping.

Intuitively, it is valuable to investigate speech data augmentation techniques applied to the phase spectrum of speech data, given the recent success in effectively utilizing phase information in voc-oders. Thus, in this paper, we propose a novel speech data augmentation method (PhasePerturbation) that employs three augmentation operations on the phase spectrum: frequency masking, temporal masking, and phase randomization. These operations draw inspiration from SpecAug \cite{chan_specaugment_2019}. We fine-tune two wav2vec2.0 pre-trained ASR models (BASE LS-960 and LARGE LV-60K) with the augmented TIMIT corpus, using different data augmentation methods. Our experimental results reveal a remarkable 10.9\% relative reduction in the word error rate (WER) compared with the baseline model fine-tuned without any augmentation operation. Additionally, the proposed method achieves further improvements when combined with the VTLP and the SpecAug over the performance of the baselines. Notably, combining our method with amplitude spectrum-based augmentations leads to even lower WER, highlighting its potential for enhancing existing amplitude-based augmentation methods.

\begin{figure}[b!]
  \centering
  \includegraphics[width=\linewidth]{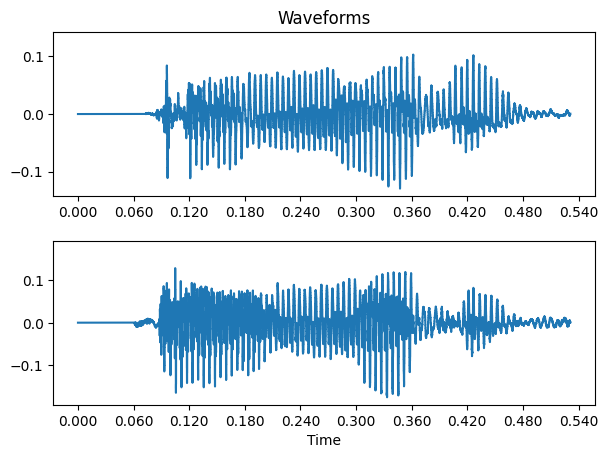}
  \includegraphics[width=\linewidth]{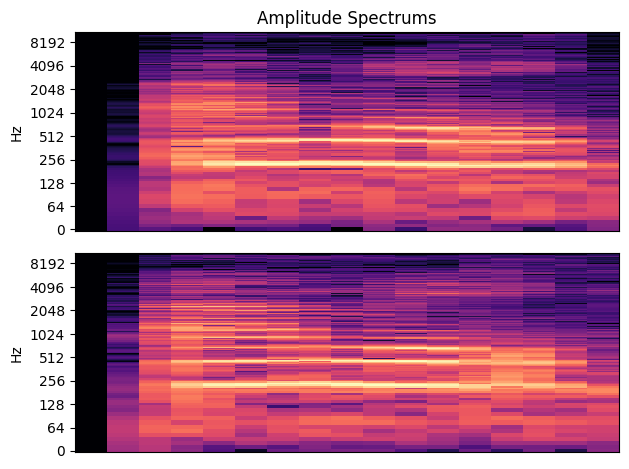}
  \centering
  \includegraphics[width=\linewidth]{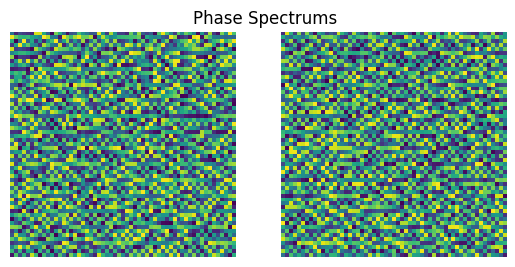}
  \caption{Waveforms (top), Amplitude spectrums (middle), and Phase spectrums (bottom) of a speaker repeating the same word twice.}
\end{figure}

\section{Background}

\subsection{Fourier Transform}

The analog speech signal is turned into a real-value discrete time sequence s[n] after sampling. Shifting the window function along the time axis to divide $ s[n] $ into time bins, these bins are sufficiently short to be treated as stationary signals, where their characteristic properties remain constant within each short time bin. The process of applying the Fourier transform to each time bin is referred to as the Short-Time Fourier Transform (STFT). STFT $ S[k,m]=F(s(n)) $ and its inverse transformation, the inversed short-time Fourier transform (iSTFT) $s(n)=F^{-1}(S[k,m])$ are represented as equation \ref{eq:eq1} and \ref{eq:eq2}:
\begin{equation}
 S[k,m]=F(s(n))=\int_{-\infty}^{+\infty} s[n]w[n-md] e^{-j2\pi k/N}  
 \label{eq:eq1}
\end{equation}

\begin{equation}
x[n]=F^{-1}(S[k,m])=\frac{1}{C} \sum_{k=0}^{N-1} \sum_{m=-\infty }^{\infty} S[k,m] e^{j2\pi k/N}
\label{eq:eq2}
\end{equation}
where $N$ is the length of STFT, and $d$ is the hop length. $k$ and $m$ are the index of the frequency bin and time bin of the time-frequency spectrum respectively. The Hanning window is represented as $w[n]=1-\cos(2\pi n/N)/2$.  $C[n]=\sum_{m=-\infty }^{\infty}w[n-md]$ is the sum, at the time $n$, of all the values of the window function moving along the time axis.

\subsection{Phase Spectrum}

The phase spectrum of a speech signal captures the phase angles associated with different frequency components. Similar to the amplitude spectrum, the phase spectrum can also be obtained through the Fourier Transform. Mathematically, the phase spectrum $\phi [k,m]$ is given by: 
\begin{equation}
  \phi [k,m]=\arctan \frac{img(S[k,m])}{real(S[k,m])}
\end{equation}
Where $\phi [k,m]$ is the phase angle of the complex representation $ S[k,m] $ at frequency $k$. The phase spectrum $\phi [k,m]$ provides information about how the different frequency components are aligned in time. In the phase spectrum, each frequency component is associated with a phase angle that indicates the timing relationship of that frequency's oscillations.

Acoustic features like MFCC and filter bank output, extracted from the amplitude spectrum, were previously preferred in speech recognition due to their connection to perceived loudness and spectral characteristics. However, their widespread use is a practical compromise driven by computational constraints. Considering the limited resources, selecting a 39-dimensional vector per frame for MFCC and an 80-dimensional vector per frame for filter bank outputs appears computationally efficient, as opposed to directly processing the raw 16 kHz speech signal, which results in a 400-sample frame vector.
Yet, with the continuous improvement in computer hardware performance and the growing prevalence of pre-trained models in speech recognition \cite{singh_novel_2023,chen_exploring_2023}, recent research has increasingly embraced the use of 16 kHz raw speech signals, incorporating phase information to attain higher accuracy. 

The phase spectrum significantly affects signal timbre, transients, and speaker distinctions. As illustrated in Figure 1, raw signal waveforms may vary, even when a speaker repeats the same sentence multiple times.

\section{Proposed Method}
\subsection{Augmentation Policy}
To introduce more diversity to the phase information, we design randomization augmentation operations for the phase spectrum. In addition, we adopt two augmentation operations on the phase spectrum (i.e., frequency masking and temporal masking) to increase the robustness of the partial loss of phase information and the partial phase loss of small speech segments. The details about the three augmentation operations are as follows.
\subsubsection{Phase randomization}
According to the equation \ref{eq:eq1}, the value of the time-frequency spectrum is a complex number that has amplitude and phase. Most previous data augmentation methods ignore the phase information of this complex number, such as MFCC and filter bank outputs. However, speech phase information contributes to the diversity of the raw speech signal in the time domain. The raw signal of a speaker repeating the same sentence several times can be of large difference because the phase of the speech signal is highly random. Inspired by the additive Gaussian white noise augmentation operation \cite{kharitonov_data_2021}, we propose phase randomization to increase the diversity of speech signals.

Initially, for phase randomization, we can assign independent random numbers to each element in the two-dimensional phase spectrum matrix of the speech signal, with dimensions $ (x,y) $, where $x$ represents the total number of frequency bins, and $y$ represents the total number of time bins. This involves generating $x\times y$ independent random numbers from a standard Normal distribution. These random numbers are then used to create a randomization matrix through element-wise multiplication with the corresponding phase spectrum matrix elements. Subsequently, the Inverse Short-Time Fourier Transform (ISTFT) can be applied to obtain the phase-randomized speech signal. However, this process introduces significant time-domain distortion, adversely affecting speech intelligibility. Karras et al. \cite{karras_training_nodate} attributed this distortion to substantial differences between adjacent frequency bins in each time frame.

To address this perceptible distortion, instead of assigning random numbers to each matrix element, we generate independent random numbers for each time bin, which are shared among elements within a column. This method requires y independent random numbers. The phase randomization operation can be summarized as Algorithm \ref{al:al1}. As demonstrated in Figure 2, our randomization method yields diverse phase-randomized speech signals while preserving nearly the same level of intelligibility as the original speech signal.

\begin{algorithm}
\DontPrintSemicolon
  \SetAlgoLined
  \KwIn {The original speech signal $s[n]$}
  \KwOut {The phase-randomized speech signal $S_{rand}[n]$}
    Apply STFT on $s[n]:S[k,m]=F(s(n))$ \;
    Calculate the phase spectrum matrix: $\phi [k,m]=argS([k,m])$ \;
    //$y$ is the column number of $\phi [k,m]$\;
    \While{$i \le y$, }{
    Sample $\mu \sim N(1,\delta ^{2})$\;
    Multiplying elements in the $i$ th th column of $\phi [k,m]$ by $\mu $ times\;
    }
    \Return $\phi [k,m]$
  \caption{Phase randomization}
  \label{al:al1}
\end{algorithm}

\begin{figure}
  \centering
  \includegraphics[width=\linewidth]{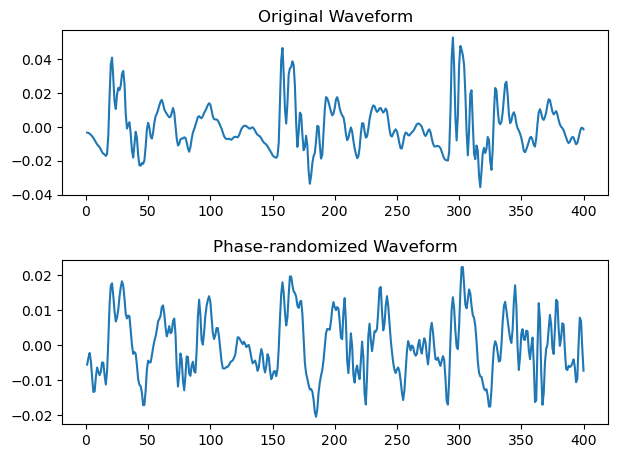}
  \caption{The original speech waveform (top) and the phase-randomized speech waveform (bottom)}
\end{figure}

\subsubsection{Frequency masking.}Frequency masking is implemented on the phase spectrum by masking a consecutive range of frequency bins $[f_{0},f_{0}+f ]$, where $f$ is initially sampled from a uniform distribution from $0$ to the maximum frequency mask width $F$, $f_{0}$ is selected from the interval $[0,v-f]$. Here, $v$ represents the total number of frequency bins in the phase spectrum.
\subsubsection{Temporal masking.}Temporal masking is implemented on the phase spectrum by masking a consecutive range of temporal bins $[t_{0},t_{0}+t ]$, where $t$ is initially sampled from a uniform distribution from $0$ to the maximum temporal mask width $T$, and $t_{0}$ is selected from the interval $[0,\tau-t]$. Here, $\tau$ represents the total number of temporal bins in the phase spectrum. Similar to SpecAug \cite{chan_specaugment_2019}, there is an upper limit imposed on the time mask to ensure it doesn't extend beyond $p$ times of $\tau$.

The phase spectrum is normalized to have a mean value of zero, thus setting the masked value to zero is effectively the same as setting it to the mean value. Figure 3 shows examples of the frequency masking and temporal masking applied to a raw speech signal. 

\begin{figure}
  \centering
  \includegraphics[width=\linewidth]{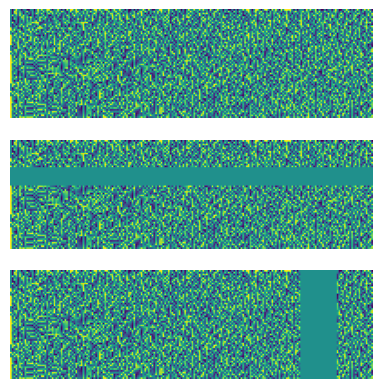}
  \caption{From top to bottom, the figures depict the phase spectrum of the original speech signal without augmentation, frequency masking, and temporal masking applied.}
\end{figure}
\section{Experiments and Results}
\subsection{Implementation Details}
One-sided, Hanning windowed STFT with a size of 1024 and a hop length of 256 is used in our experiment. Phase augmentation parameters are summarized in Table 1. $F$ and $T$ represent the maximum widths of the frequency mask and the temporal mask, respectively. We consider policies where multiple frequency and time masks are applied. $m_{F}$ and $m_{T}$ denote the number of frequency and time masks applied. The multiple masks may overlap. $p$ represents the maximum ratio of the temporal mask width to the speech length.

\begin{table}[b]
  \caption{The masking parameters.}
  \label{tab:tab1}
  \begin{tabular}{ccccc}
    \toprule
    $F$ & $m_{F}$ & $T$ & $m_{T}$ & $p$ \\
    \midrule
    10 & 2 & 45 & 2 & 0.1\\
  \bottomrule
\end{tabular}
\end{table}

\subsection{Model}
For our experiments, we adopt the self-supervised Wav2vec2.0 \cite{baevski_wav2vec_2020} as the backbone for the ASR model. The Wav2vec2.0 model was pre-trained with large amount of unlabeled data in a self-supervised manner. We fine-tune the Wav2vec2.0 model with the augmented TIMIT corpus, using our proposed data augmentation policies. To verify the performance of our phase augmentation method on models with different sizes of parameters, two variants of wav2vec2.0 models are used: BASE LS-960 and LARGE LV-60K. BASE and LAR-GE models differ in Transformer architecture: The BASE model consists of 95 million parameters, featuring 12 transformer blocks, a model dimension of 768, an inner dimension (FFN) of 3072, and 8 attention heads. In contrast, the LARGE model is composed of 315 million parameters, incorporating 24 transformer blocks, a model dimension of 1024, an inner dimension of 4096, and 16 attention heads. The BASE LS-960 variant was pre-trained on LibriSpeech 960h \cite{panayotov_librispeech_2015}, while the LARGE LV-60K model underwent pre-training on 60k hours of unlabeled speech data from LibriVox \cite{kahn_libri-light_2020}. To reduce overfitting, the time-step and channel masking are applied on the outputs of the encoder of wav2vec2.0 model as implemented in the original wav2vec2.0 paper \cite{baevski_wav2vec_2020}.

Both pre-trained models are fine-tuned for 100 epochs with Adam optimizer with a learning rate of $3e^{-5}$. The TIMIT dataset consists of default TRAIN and TEST sets. To fine-tune the pre-trained wav2vec 2.0 models, we use a 90\% TRAIN set to train the model and a 10\% TRAIN set for validation.

When fine-tuning the wav2vec2.0 models, 7 augmentation methods are applied to the raw waveform speech data: (1) the original TIMIT dataset without augmentation, (2) PhaseAug \cite{lee_phaseaug_2023}, (3) Vocal Tract Length Perturbation (VTLP) \cite{jaitly_vocal_2013}, (4) proposed PhasePerturbation, (5) the combination of PhasePerturbation and VTLP, (6) SpecAug \cite{chan_specaugment_2019}, and (7) the combination of PhasePerturbation and SpecAug. The VTLP and the SpecAug are both applied on the amplitude spectrum and then use iSTFT to transform them back to the time domain.
\begin{table}[b]
  \caption{WER (\%) of BASE LS-960 and LARGE LV-60K with five augmentation methods on the TIMIT.}
  \label{tab:tab2}
  \begin{tabular}{clc}
    \toprule
    Models & Policies & WER (\%) \\
    \midrule
    \multirow{7}{*}{BASE LS-960} & None & 21.6 \\
    & PhaseAug[1] & 21.2 \\
    & VTLP[2] & 20.2 \\
    & PhasePerturbation & 19.4 \\
    & PhasePerturbation+VTLP & 19.1 \\
    & SpecAug[3] & 18.9 \\
    & PhasePerturbation+SpecAug & \textbf{18.5} \\
    \midrule
    \multirow{7}{*}{LARGE LV-60K} & None & 20.1 \\
    & PhaseAug[1] & 19.7 \\
    & VTLP[2] & 18.8 \\
    & PhasePerturbation & 17.9 \\
    & PhasePerturbation+VTLP & 17.5 \\
    & SpecAug[3] & 17.4 \\
    & PhasePerturbation+SpecAug & \textbf{16.9} \\
  \bottomrule
\end{tabular}
\end{table}
\subsection{Experimental Results}
In Table \ref{tab:tab2}, our phase augmentation technique shows excellent performance across all models, yielding 19.4\% and 17.9\% WER for the BASE LS-960 model and the LARGE LV-60K model, respectively. The relative gains our method offers over the original TIMIT dataset, are about 10.1\% and 10.9\% across both the Wave2vec2.0 models. Our method outperforms the static PhaseAug approach by 1.8\% WER. Moreover, our technique surpasses VTLP's performance on both the BASE LS-960 model and the LARGE LV-60K model. It is worth highlighting that our dynamic phase augmentation method attains even lower WER (17.5\% and 16.9\%) when combined with the VTLP and the SpecAug, which are both amplitude spectrum augmentation operations. The results demonstrate the effectiveness of our method for complementing existing augmentation methods that operate on the amplitude spectrum.

\section{Conclusion}
In this paper, we propose a novel dynamic phase spectrum augmentation method for speech data, PhasePerturbation. Our method boosts augmentation diversity by employing three phase spectrum operations: phase randomization, frequency masking, and temporal masking. The experimental results demonstrate the efficacy of our phase-based augmentation method in improving speech recognition performance, especially when dealing with limited datasets. To validate our method, we compare it to VTLP and PhaseAug, revealing that our dynamic phase augmentation outperforms static phase operations. Our experiments also demonstrate the potential of combining amplitude spectrum-based methods with our phase spectrum-based method to enhance data diversity.

It is noteworthy that our phase augmentation technique impacts only the phase spectrum, not the amplitude spectrum, like the Mel-spectrum. This limitation restricts its usage to ASR systems that take raw waveform speech data as input. However, this distinction sets our method apart, as many existing augmentation techniques primarily manipulate the amplitude spectrum, as seen in methods such as SpecAug++ \cite{wang_specaugment_2021} or CyclicAugment \cite{wang_cyclicaugment_2022}. In our future research, we plan to explore the integration of phase augmentation operations with state-of-the-art amplitude spectrum methods.
\begin{acks}
This work is supported by the 2020 Catalyst: Strategic New Zealand - Singapore Data Science Research Programme Fund by the Ministry of Business, Innovation and Employment (MBIE), New Zealand.
\end{acks}

\bibliographystyle{ACM-Reference-Format}
\bibliography{paper_citation}

\end{document}